\documentclass[preprint2]{aastex7}

\def\bfnote{}

\def\cm3{cm$^{-3}$}
\def\apj{ApJ}
\def\apjl{ApJL}

\def\aap{A\&A}

\usepackage{graphicx}
\usepackage{epsf}
\usepackage{natbib}
\usepackage{hyperref}
\usepackage{lipsum}

\newcommand{\simlt}{\lower.5ex\hbox{$\; \buildrel < \over \sim \;$}}

\begin{document}

\title{Composition of planetary debris around the white dwarf GD 362}

\shorttitle{Mid-infrared spectrum of GD 362}

\author[0000-0001-8362-4094]{William T. Reach}
\affil{Space Science Institute, 4765 Walnut St, Suite B, Boulder, CO 80301, USA}
\email{wreach@spacesciece.org}

\author[0000-0001-6098-2235]{Mukremin Kilic}
\affil{Homer L. Dodge Department of Physics and Astronomy, University of Oklahoma, 440 W. Brooks Street, Norman, OK 73019, USA}
\email{mukreminkilic@gmail.com}

\author[0000-0002-9548-1526]{Carey M. Lisse}
\affil{Planetary Exploration Group, Space Department, Johns Hopkins University Applied Physics Laboratory, 11100 Johns Hopkins Road, Laurel, MD 20723, USA}
\email{Carey.Lisse@jhuapl.edu}

\author[0000-0002-1783-8817]{John H. Debes}
\affil{Space Telescope Science Institute, 3700 San Martin Dr., Baltimore, MD 21218, USA}
\email{debes@stsci.edu}

\author[0000-0002-5775-2866]{Ted von Hippel}
\affil{Department of Physical Sciences and SARA, Embry-Riddle Aeronautical University, Daytona Beach, FL 32114, USA}
\email{ted.vonhippel@erau.edu}

\author[0000-0002-7898-6194]{Bianca Azartash-Namin}
\affil{Homer L. Dodge Department of Physics and Astronomy, University of Oklahoma, 440 W. Brooks Street, Norman, OK 73019, USA}
\email{Bianca.Azartash.Namin-1@ou.edu}

\author[0000-0003-0475-9375]{Lo\"ic Albert}
\affil{D\'epartement de Physique, Observatoire du Mont-M\'egantic and Trottier Institute for Research on Exoplanets, Universit\'e de Montr\'eal, C.P. 6128, Succ. Centre-ville, Montr\'eal, H3C 3J7, Québec, Canada}
\email{loic.albert@umontreal.ca}

\author[0000-0001-7106-4683]{Susan E. Mullally}
\affil{The Space Telescope Science Institute, 3700 San Martin Dr., Baltimore, MD 21218, USA}
\email{smullally@stsci.edu}

\author[0009-0004-7656-2402]{Fergal Mullally}
\affil{Constellation, 1310 Point Street, Baltimore, MD 21231, USA}
\email{fergal.mullally@gmail.com}

\author[0000-0002-7698-3002]{Misty Cracraft}
\affiliation{The Space Telescope Science Institute, 3700 San Martin Dr., Baltimore, MD 21218, USA}
\email{cracraft@stsci.edu}

\author[0009-0001-7595-5704]{Madison Bernice}
\affil{Homer L. Dodge Department of Physics and Astronomy, University of Oklahoma, 440 W. Brooks Street, Norman, OK 73019, USA}
\email{mrb078@latech.edu}

\author[]{Selin L. Erickson}
\affil{Homer L. Dodge Department of Physics and Astronomy, University of Oklahoma, 440 W. Brooks Street, Norman, OK 73019, USA}
\email{selinlily@ou.edu}

\begin{abstract}

White dwarf stars with high abundances of heavy elements in their atmospheres
and infrared excesses are believed to be accreting planetary material.
GD 362 is one of the most heavily polluted white dwarfs and has an exceptionally strong mid-infrared excess, reprocessing 2.4\% of the star's light into the mid-infrared.
We present a high signal-to-noise, medium-resolution spectrum of GD 362
obtained with JWST, covering 0.6--17 $\mu$m, along with photometry out to 25.5 $\mu$m.
The mid-infrared spectrum is dominated by an exceptionally strong 9--11 $\mu$m silicate feature, which can be explained by a combination of 
olivine and pyroxene silicate minerals.
{\bfnote Grains such as carbon, hotter than silicates, are required to explain the near-infrared emission.}
The silicates and carbon reside in a disk from 
140 to 1400 stellar radii, and the disk scale
height is greater than half the stellar radius.
The elemental abundances of the solid material, relative to Si, are
within a factor of 2 of meteoritic (CI chondrites) for C, O, Mg, Al, and Fe,
with Al elevated and O slightly depleted. A similar pattern is observed for
the abundances of accreted material in the stellar photosphere.
Hydrogen is an exception, because no significant H-bearing minerals or water
were detected in the disk, despite a large H abundance in the photosphere.
\end{abstract}

\section{Introduction}

Infrared spectroscopy of stellar remnants provides a unique opportunity to measure
the detailed composition of solid material from other planetary systems.
Most stars in the solar neighborhood 
will evolve into white dwarfs. The fate of planetary
systems during post-main sequence evolution is an important question, and
the remnant white dwarfs may host long-lived habitable zones
\citep{agolTransitSurveysEarths2011,vanderburgLonglivedHabitableZones2025}. \citet{villaverCanPlanetsSurvive2007} studied the survivability of gas giants, and found that small planets
within a few au of the star should be destroyed during the planetary nebula phase \citep[also see][]{schroderDistantFutureSun2008}, while the outer planets should survive. 

Several white dwarfs  host giant planet candidates \citep{mullallyJWSTDirectlyImages2024,limbachMIRIExoplanetsOrbiting2024,
limbachThermalEmissionConfirmation2025}.
The candidate planets closer to their star (and not spatially resolved from it) fall within the `forbidden zone' where planets would be destroyed during the host star's giant phase;  their existence points to a reshuffling of planetary systems during stellar evolution. 

\citet{debesAreThereUnstable2002} presented a dynamical instability model that can populate the inner 5 au around a white dwarf with planetesimals.
Mass loss during a star's post-main-sequence evolution
can drive planetesimal orbits into interior mean-motion resonances
with a giant planet. These planetesimals are slowly removed through chaotic excursions of eccentricity that create radial orbits capable of tidally disrupting the planetesimals and forming debris disks \citep{debesLinkPlanetarySystems2012,verasFormationPlanetaryDebris2015}. We now recognize approximately 50 such dust disks around white dwarfs, including
approximately 20 systems with a gas disks \citep{becklinDustyDiskGD2005,vonhippelNewClassDusty2007,farihiInfraredSignaturesDisrupted2009,barberFrequencyDebrisDisks2012,rocchettoFrequencyInfraredBrightness2015,dennihyFiveNewPostmainsequence2020,melisSerendipitousDiscoveryNine2020,gentilefusilloWhiteDwarfsPlanetary2021}, corresponding to a 
fraction of $\sim1.5$\%
of white dwarfs having detectable dusty disks
 \citep{wilsonUnbiasedFrequencyPlanetary2019}.

Dusty white dwarfs provide a unique opportunity to study the composition of exoplanetary material through studies of the photospheric abundances
of the accreted material \citep[e.g.][]{zuckermanChemicalCompositionExtrasolar2007} or disk mineralogy \citep{reachDustCloudWhite2005,reachDustCloudWhite2009,balleringGeometryG2938White2022}. Even though
we know thousands of planets around other stars, elemental abundance measurements of planetary
material in and around white dwarfs is the only method that provides a bottom-up view
of bulk composition. There are only two dusty white dwarfs that were sufficiently bright to be studied spectroscopically with Spitzer, G29-38 and
GD 362. \citet{juraSixWhiteDwarfs2009} observed six additional white dwarfs with Spitzer, but with a S/N $\sim3$-5 those spectra did not provide meaningful
constraints on the disk composition. 
Among the eight dusty white dwarfs with Spitzer spectra, GD 362 is notable for its exceptionally
strong infrared excess and metal abundance in its atmosphere.

GD 362 has a helium-dominated atmosphere, but it possesses an anomalously large mass of hydrogen 
\citep[][$\sim0.01~M_{\earth}$]{juraXRayInfraredObservations2009}.
GD 362 is also accreting
at least 16 different elements from its circumstellar disk \citep{zuckermanChemicalCompositionExtrasolar2007}. An intriguing possibility is that
it might have accreted its atmospheric hydrogen from one or multiple parent bodies with internal ice \citep[e.g.,][]{hoskinWhiteDwarfPollution2020}. \citet{malamudPostmainSequenceEvolution2016} showed that a significant fraction of the total water content of a rocky object is retained throughout the different stellar evolution phases.
Based on measurements of atmospheric abundances, infrared excess, and X-ray limits, \citet{juraXRayInfraredObservations2009} suggested that
the large mass of hydrogen in GD 362 could be explained by accretion of a large swarm of water-rich asteroids or a parent body analogous to Callisto.

In order to constrain the mineralogy of its disk and to search for unambiguous signs of water in GD 362, we obtained a high signal-to-noise ratio,
medium resolution spectrum with  JWST.
This paper is organized as follows. After summarizing properties of GD 362, the observations and data analysis are described in \S\ref{sec:obs}.
Then we turn to modeling the spectrum using two independent techniques. In \S\ref{sec:model}, we use a radiative transfer code with an axisymmetric, parameterized disk to constrain the distribution of dust around the star.
In \S\ref{sec:mineralogy}, we make a linear decomposition of the spectrum using a library of minerals to 
constrain the mineralogy of the dust.
We then discuss some of the results (\S\ref{sec:discussion}) and summarize the conclusions (\S\ref{sec:conclusion}).

\section{Properties of GD 362}
The target of this study is GD 362 (WD 1729+371), a DAZB (metal-enriched with primarily helium atmosphere) white dwarf. Basic properties
of the star are listed in Table~\ref{tab:gd362}, largely from the Montr\'eal White Dwarf Database\footnote{www.montrealwhitedwarfdatabase.org} 
%\href{https://www.montrealwhitedwarfdatabase.org/WDs/GD%20362/GD%20362.html}{www.montrealwhitedwarfdatabase.org}} 
\citep{dufourMontrealWhiteDwarf2017}.
Modern estimates of the white dwarf properties are much different from the 
initial ones, which had a mass of 1.2 $M_\odot$ which would have made 
it the most massive
metal-polluted white dwarf \citep{gianninasDiscoveryCoolMassive2004}. 
The distance has in the meantime been updated using parallax including ground-based \citep{kilicDirectDistanceMeasurement2008} and space-based (Gaia DR3).
The primary reason for the difference in the derived mass is a change in the surface gravity determination,
initially based on a hydrogen-dominated atmosphere and updated
when it was found to be 
helium dominated \citep{zuckermanChemicalCompositionExtrasolar2007}.
%Table~\ref{tab:gd362} summarizes the properties of the white dwarf used in this paper.

\begin{figure}
\begin{center}
\includegraphics[width=.3\textwidth]{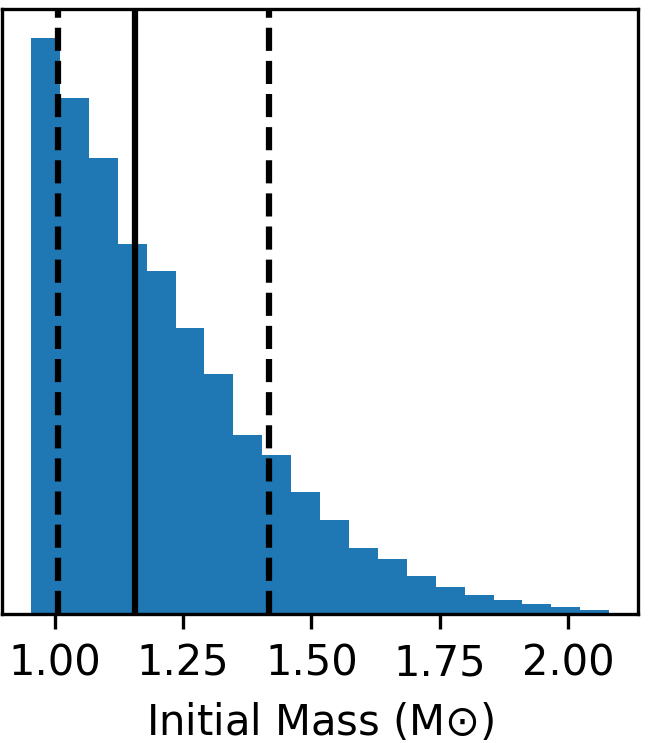}
\vskip -.4in
\end{center}
\caption{Likelihood distribution for the initial mass of GD 362. The maximum likelihood initial-mass is $1.24\pm 0.13$ $M_\odot$
\label{fig:mi}}
\end{figure}

\begin{table}
  \caption{Properties of White Dwarf GD 362 \label{tab:gd362}} 
  \begin{center}      
  \begin{tabular}{lll}
\hline
\multicolumn{2}{c}{Basic Properties}\\ \hline
J2000 Coordinates  & 17~31~34.35, +37~05~17.24 & DR3\\
Distance & 56.1 pc &  DR3\\
Proper motion & (24.1, -217.0) mas yr$^{-1}$ &  DR3\\
$T_{\rm eff}$ & $9825\pm 59$ K & C23\\
$\log{g}$ & $7.99\pm 0.01$ cm~s$^{-2}$ & C23\\
Mass & $0.574\pm0.007$ $M_{\odot}$ & C23\\
Luminosity & 0.0016 $L_{\odot}$ & G12\\
\hline \multicolumn{2}{c}{Derived Properties}\\ \hline
WD Radius & $8790\pm 180$ km & \\
Initial mass      & $1.16_{-0.15}^{+0.26}$ $M_\odot$ & C18\\ 
%Main sequence age & $6.9_{-3.39}^{+4.31} Gyr & CD16\\
WD cooling age    & $0.68\pm 0.01$ Gyr\ & K22\\
Total age         & $>4.2$ Gyr & K22\\
\hline
  \end{tabular}
  \end{center}
%  \begin{center}
{References}: DR3 \citep{gaiacollaborationGaiaDataRelease2023,gentilefusilloCatalogueWhiteDwarfs2021}, C23=\citep{caronSpectrophotometricAnalysisCool2023}, G12 \citep{giammicheleKnowYourNeighborhood2012},
C18 \citep{cummingsWhiteDwarfInitial2018}, CD16 \citep{choiMESAISOCHRONESStelLAR2016,dotterMESAISOCHRONESStelLAR2016},
K22 \citep{kimanWdwarfdatePythonPackage2022}
%  \end{center}
\end{table}

\begin{figure*}
\centering
\includegraphics[width=\textwidth]{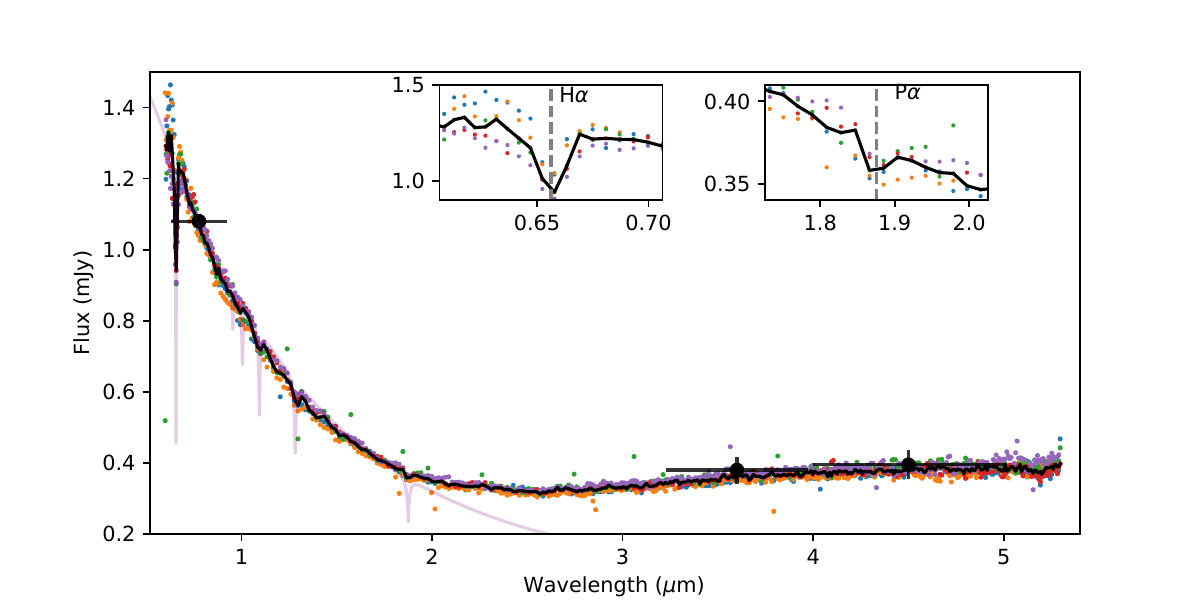}
\caption{The JWST/NIRSPEC spectrum of GD 362, covering 0.6--5.3 $\mu$m. 
Dots show spectra from each exposure, with the color changing for each of the 5 nods along the slit.
Insets show the spectra in the vicinity of the H$\alpha$ and P$\alpha$ photospheric lines of the white dwarf.
The black curve shows the median of the 5 spectra. The black dots show the Spitzer/IRAC and Gaia photometry. 
For comparison, a white dwarf photosphere model, normalized to the Gaia $G_{\rm RP}$ flux, is shown in light purple. \\
 \label{fig:nirspec}}
\end{figure*}

The cooling age and likely main sequence progenitor were determined from the Bayesian procedure
{\tt wdwarfdate} \citep{kimanWdwarfdatePythonPackage2022},
using an initial-final mass relation  and stellar
evolution isochrones.
The effective temperature and surface gravity are from \citet{caronSpectrophotometricAnalysisCool2023}, who
fit the GD362 spectral energy distribution 
(with Gaia parallax and Pan-STARRS photometry)
using a He atmosphere model with trace amounts of H and Ca (H/He=0.03, Ca/He=$10^{-7}$)
based on observations by \citet{zuckermanChemicalCompositionExtrasolar2007}.
We imposed an {\it a priori} lower limit on the initial mass of 0.95 $M_\odot$, so that the star could become a
white dwarf within the age of the universe.
%The confidence intervals for the
%inferred parameters are at the 33$^{rd}$ and 67$^{th}$ percentiles, corresponding to
%$\pm 1\sigma$.
The maximum likelihood estimate of the initial mass is somewhat larger than the Sun,
with a spectral type in the range F8--F2 when it was on the main sequence; 
Figure~\ref{fig:mi} shows the likelihood distribution for the initial mass. 
The progenitor mass remains somewhat dependent upon the way the white dwarf photometry is fitted and the initial-to-final mass relation.

The white dwarf remains relatively hot, as it has only evolved off the main sequence for 0.7 Gyr.
The total age of the star, including main sequence and cooling time, is
a fiducial age estimate for the properties of planets that formed contemporaneously with the star. 
Because giant planets are self-luminous, their brightness depends on this total age; 
at the maximum likelihood estimate of 7.6 Gyr, this means any giant planets would be comparable to or somewhat cooler than those of the Solar System.
Experimenting with different white dwarf photometry to determine how
well the age is constrained, we found a total age as short as 3.5 Gyr 
(A0 main sequence progenitor) is ruled out by the 
observations. The Bayesian confidence interval is $>4.2$ Gyr.

\section{Observations and data analysis\label{sec:obs}}
\begin{figure*}
\centering
\includegraphics[width=\textwidth]{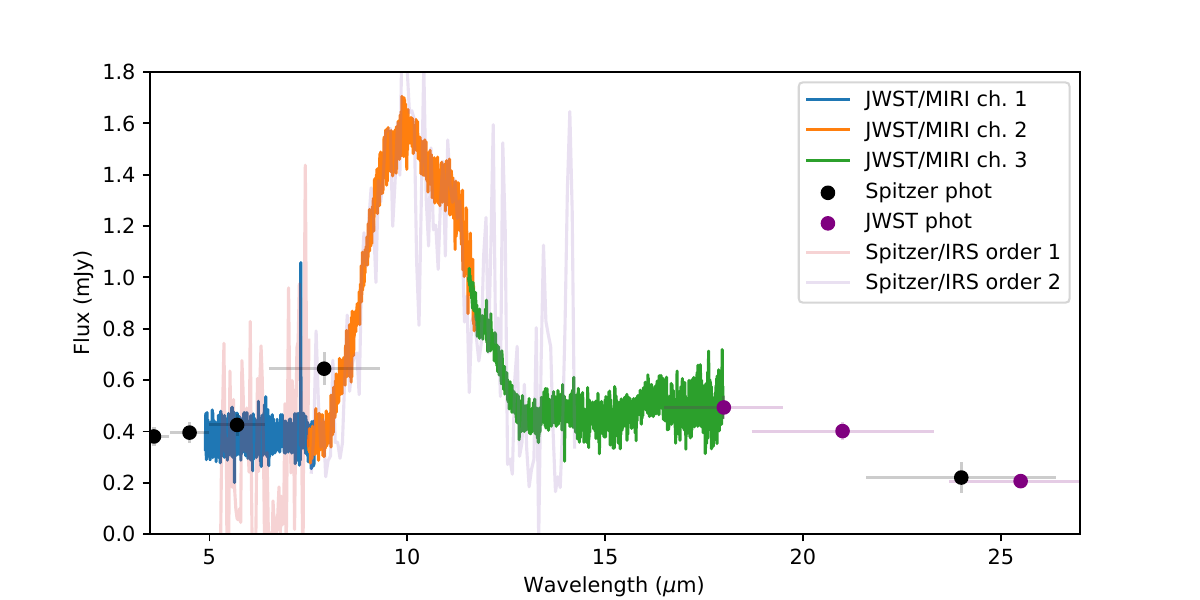} 
\caption{The JWST/MIRI spectrum of GD 362. The three colored curves show the spectra in MIRI/MRS channels 1, 2, and 3. The channel 4 spectrum is not shown because the star was not detected. The single-channel spike at 7.311 $\mu$m is 
likely an artifact. Black points show the
photometry in five bands from Spitzer \citep{juraInfraredEmissionDusty2007}, and purple dots show the 
photometry in three bands from JWST. 
 \label{fig:mrs}}
\end{figure*}

\subsection{JWST/NIRSPEC Observations}

The near-infrared spectrum was obtained with the Near InfraRed Spectrograph 
\citep[NIRSpec][]{jakobsenNearInfraredSpectrographNIRSpec2022}, through the S200A1 slit ($0.2''\times 3.3''$) and prism 
(0.6--5.3 $\mu$m at resolving power 30--330) as part of the GO program 2919 (PI M. Kilic). Five exposures of 26 sec were taken with small dithers along
the slit, for a total exposure of 132 sec. The observations were performed on 2024 Jun 28.
The data were reduced with the JWST calibration pipeline version 1.15.1 \citep{bushouseJWSTCalibrationPipeline2024}. Both the raw and fully reduced data were retrieved from the Mikulski Archive for Space Telescopes (MAST). These data can be accessed via \dataset[doi: 10.17909/cyd9-ga24]{https://doi.org/10.17909/cyd9-ga24}.

The extracted spectrum of GD 362 for each of the 5 exposures is shown in Figure~\ref{fig:nirspec}.
At wavelengths shorter than 2.5 $\mu$m, the observed flux is dominated by direct flux from the star.
Two Balmer absorption features from the photosphere are prominent despite the low spectral resolution.
H$\alpha$ and P$\alpha$ are shown in the insets of Fig.~\ref{fig:nirspec}, and P$\beta$ and P$\gamma$ are also present.
At wavelengths longer than 2.5 $\mu$m, the spectrum rises even though the white dwarf photosphere decreases
as the inverse square of wavelength in its Rayleigh-Jeans spectrum. The emission at these wavelengths
is  dominated by circumstellar material.

To better separate the star from the circumstellar material, we scaled a model white dwarf photosphere 
to the optical photometry. 
The photosphere model is for a 10,000 K effective temperature white dwarf with $\log g$=8
\citep{koesterWhiteDwarfSpectra2010}, close to the properties of GD 362 in Table~\ref{tab:gd362}. 
{\bfnote Although GD 362 is a DB white dwarf, we show the model is for a DA white dwarf, which includes H absorption lines, deeper than those of GD 362.}
We normalized the star to the Gaia DR3 photometry in
the `red part' of its passband, where the catalog flux is 1.08 mJy at 0.777 $\mu$m,
then scaled downward by an additional -3\% to better match the JWST spectrum across the near-infrared where
the star dominates over the disk emission.

% does scattered light contribute to the near-infrared? can't really separate from photosphere easily
% but can test idea using mcfost models

\subsection{JWST/MIRI Observations}

The mid-infrared spectrum was obtained with the Mid-InfraRed Instrument 
\citep[MIRI][]{wrightMidinfraredInstrumentJWST2023},
using the Medium Resolution Spectrograph (MRS). Twelve exposures were taken, comprising four dithers in each of the
short, medium, and long spectral settings. For the short and medium settings, exposure times were 979 sec per dither; for the long spectral setting, the exposure times were 1362 sec per dither. The combined spectrum covers wavelengths 5--28 $\mu$m.
The longest-wavelength channel 4 (wavelengths $>17.7$ $\mu$m) data had low signal-to-noise, as expected
at the time of planning the observations, and they are not used; they came `for free' with the channel 1 observations.

To extend the spectral energy distribution beyond the usable part of the spectrograph, images were obtained with MIRI in the F1800W, F2100W, and F2550W filters, which are centered at 18, 21, and 25.5 $\mu$m,
respectively. The number of dithers $\times$ exposure time per dither in the three filters was
$4\times 36$~sec, $4\times 58$ sec, and $17\times 80$~sec, respectively.
All observations were performed on 2024 Jun 28, the same date as the NIRSPEC observation.

The extracted spectrum of GD 362, and the filter photometry, are shown in Figure~\ref{fig:mrs}.
At the shortest MIRI wavelengths (mostly in channel 1), the spectrum is flat, but the spectrum
rises rapidly in a tremendously strong 9--11 $\mu$m feature that has an amplitude four times that of the relatively flat continuum.
At wavelengths longer than the feature, the continuum slowly rises until the longest wavelength of the MRS
spectrum (18 $\mu$m). The filter photometry agrees very well with the flux extracted from the spectrum.
After this wavelength the spectrum declines. 

{\bfnote The MIRI/MRS and NIRSPEC spectra agree astonishingly well at their overlapping wavelengths near 5 $\mu$m, allowing them to be readily merged.}

\subsection{Luminosity of the disk}

Using the spectral energy distribution from the near and mid-infrared wavelengths, we can 
calculate the total luminosity of the disk, 
because the peak in the star's flux lies within the JWST-observed range.
For this purpose, we approximate the spectral energy distribution as the sum of two blackbodies (temperatures 826 K and 305 K)
for the continuum, plus a Gaussian for the 10 $\mu$m silicate feature. 
The integrated luminosity is $3.0\times 10^{-5} L_\odot$,
of which 22\% arises from the 10 $\mu$m silicate feature.
The disk luminosity is a remarkably large 2.4\% of the  star's luminosity, reprocessed as infrared radiation by the 
circumstellar dust.

\subsection{Stability of dust emission from GD 362}

The JWST/MIRI and Spitzer data are compared in Figure~\ref{fig:mrs}.
The Spitzer spectrum is the IRS enhanced
data product\footnote{\href{https://doi.org/10.26131/irsa487}{DOI:10.26131/irsa487}}  from the Spitzer Heritage Archive
\citep{irsinstrumentteamandscienceusersupportteamIRSInstrumentHandbook2012}.
Photometry at 3.6, 4.5, 5.8, 8, and 24 $\mu$m is from \citet{juraInfraredEmissionDusty2007},
with small additional corrections, to account for the actual shape of the GD 362 spectrum within the filter bandpasses,  of -0.5\% at 8 $\mu$m and -2.7\% at 24 $\mu$m \citep{iracinstrumentandinstrumentsupportteamsIRACInstrumentHandbook2021,mipsinstrumentandmipsinstrumentsupportteamsMIPSInstrumentHandbook2011}.
Starting from the longest overlapping wavelengths:
the Spitzer 24 $\mu$m flux was 6.7\% higher than that from  JWST at 25.5 $\mu$m.
This difference is partly due to the spectral energy distribution, taking into account
the different filter bandpasses (Fig.~\ref{fig:mrs}).
The Spitzer 8 $\mu$m band includes the steeply rising silicate feature; it appears consistent to first order with the MIRI
channel 2 MRS spectrum. 
The Spitzer/IRS spectrum approximately matches the JWST/MIRI one. 
Specifically,
the amplitude and width of the 9--11 $\mu$m silicate feature are consistent. The Spitzer spectrum dips lower than that from JWST in a small range from 6.5--6.8 $\mu$m that is likely an artifact of the
Spitzer spectrum; otherwise, the levels of the two spectra agree reasonably well (to within 10\%).

The Spitzer 5.8 $\mu$m flux is 8\% higher than the MIRI channel 1 MRS spectrum at overlapping wavelengths. 
The 3.6 and 4.5 $\mu$m Spitzer bands overlap the NIRSPEC spectrum, and they are 5\% and 3\%, respectively,
brighter than the corresponding part of the JWST spectrum. The spectrum is flat and featureless over
this wavelength range, and no color corrections are required for the IRAC bandpasses.

Both the photometry and spectra obtained with JWST and Spitzer nearly agree flux level and shape,
indicating that the composition of dust around GD 362 did not  change much
over the 18-year interval between observations with the two spacecraft.
The differences between Spitzer and JWST  are consistent at multiple wavelengths, in the
sense that the Spitzer observations were higher by 3--7\%. 
The absolute calibration of
Spitzer/IRAC at 3.6, 4.5, 5.8, and 8 $\mu$m is 2\% \citep{reachAbsoluteCalibrationInfrared2005}.
JWST/NIRSPEC and MIRI have spectroscopic calibration requirements of 10\% and 15\%, respectively,
but in-flight data show considerably better performance.
The JWST/MIRI MRS cross-calibration shows that MRS is consistent with MIRI imaging and Spitzer to
within 1--3\% \citep{lawJamesWebbSpace2025}. For GD 362, this means that the dust brightness was $5\pm 3$\% brighter in 2006
than in 2024. 
%There is a hint of a trend, such that the longer-wavelength observations ($>5.8$ $\mu$m) changed more than the shorter wavelengths.

Thus it appears there has been no significant new dust production (or loss) over the interval between
observations, while the amount of dust may have slightly decreased.
The similarity of the spectra shows that the different instrumental artifacts (e.g. the Spitzer/IRS 14 $\mu$m `teardrop' and the JWST `12 $\mu$m blue object leak') do not contaminate the spectrum, allowing us to  
confidently distinguish systematics from real dust emissivity features.

\begin{figure}
\begin{center}
\includegraphics[width=.472\textwidth]{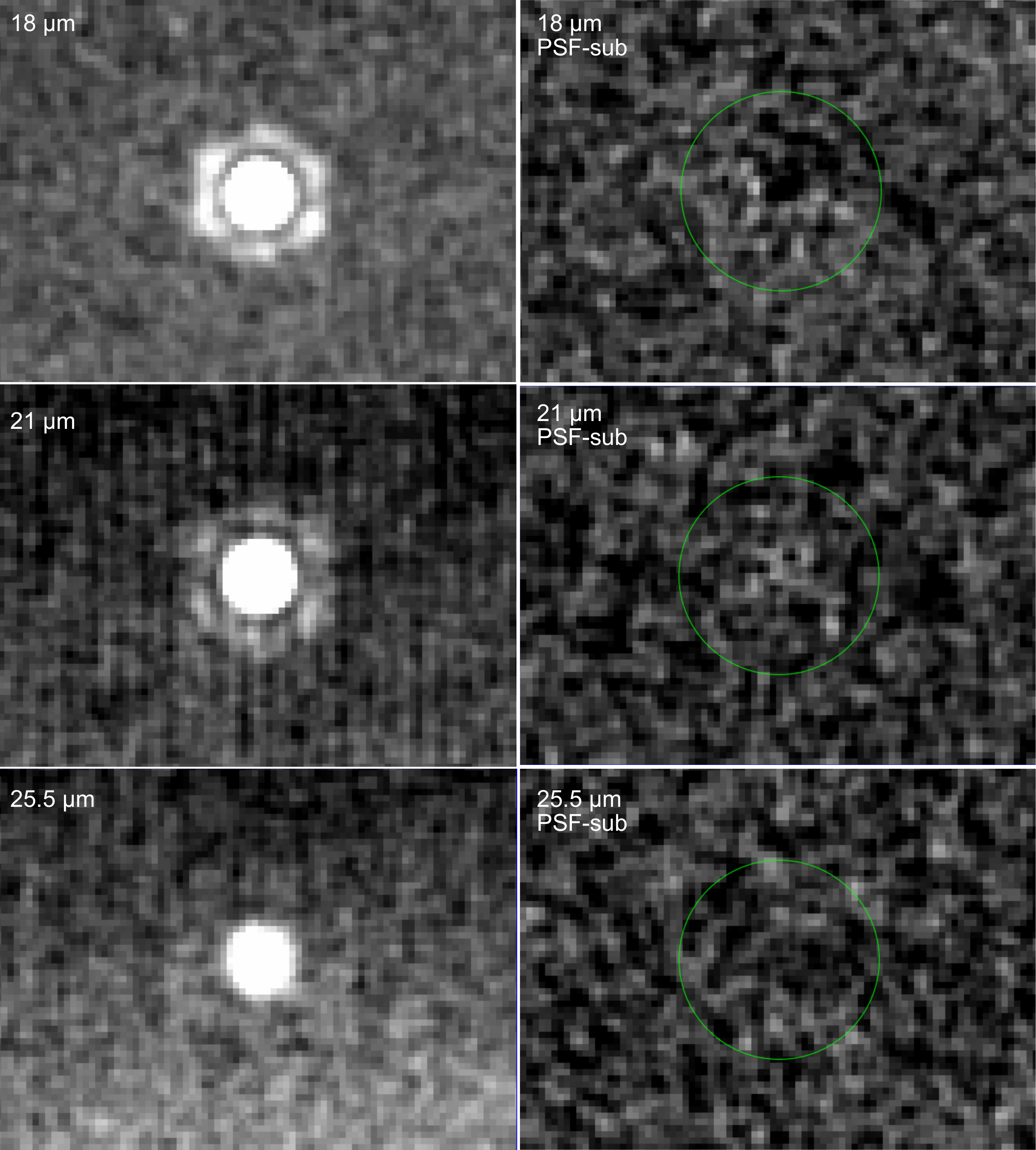}
\end{center}
\caption{Images of GD 362 in three JWST/MIRI filters at 18 $\mu$m ({\it top}), 21 $\mu$m ({\it middle}), and 25.5 $\mu$m ({\it bottom}).
For each filter, the direct images are shown on the ({\it left}), and the images after subtraction of a point spread function centered on the star are
in the ({\it right}). In the PSF-subtracted images, a green circle denotes 100 au from the central star.
\label{fig:sixpanel}}
\end{figure}

\section{Exoplanet search}
The JWST 18--25.5 $\mu$m images of GD 362 are the deepest ever made for this star, and thus they allow the best opportunity yet to 
directly detect resolved exoplanet companions. Because of the low luminosity of the white dwarf and the self-luminosity of giant planets, large planets 
can be directly detected without the need for a coronagraph or other techniques to mask the host star. 
The technique is described in \citet{poulsenMIRISearchPlanets2024} and \citet{mullallyJWSTDirectlyImages2024}.
GD 362 is toward the outer limit of distance for this technique, because even JWST's angular resolution (FWHM) of $0.59''$ at 18 $\mu$m \citep{wrightMidinfraredInstrumentJWST2023}
corresponds to a physical scale (radius) of 17 au.
The first, bright Airy ring of the point spread function is 23 au away from the star. 
Though the white dwarf has an intrinsically low luminosity, the point spread function (PSF) still limits any planet search within 30 au radius. 
To mitigate this, a PSF was derived from a star in the same image. This way, most instrumental artifacts (in particular, the orientation of the telescope 
structure with respect to the detector) are identical. A shifted image of the PSF star was subtracted from the portion of the image containing the white dwarf.

\def\tnm{\tablenotemark}
\begin{deluxetable}{ccccc}
\tablecolumns{5}
\tablecaption{Materials\label{tab:materials}} 
\tablehead{
\colhead{Name} & \colhead{Material} & \colhead{Composition} & \colhead{Reference} & \colhead{}
}
\startdata
DrSil    & Draine silicates & unknown & \citet{draineOpticalPropertiesInterstellar1984}\\
AmCar    & amorphous carbon & C & \citet{edohOpticalPropertiesCarbon1983}\\
crsil    & crystalline silicates & & \citet{liUltrasmallSilicateGrains2001} \\
olmg100  & amorphous olivine & Mg$_2$SiO$_4$ & \citet{dorschnerStepsInterstellarSilicate1995a}\\
olmg50   & amorphous olivine & MgFeSiO$_4$ & \citet{dorschnerStepsInterstellarSilicate1995a}\\
pyrmg100 & amorphous pyroxene & MgSiO$_3$ & \citet{dorschnerStepsInterstellarSilicate1995a}\\
pyrmg50  & amorphous pyroxene & Mg$_{0.5}$Si$_{0.5}$O$_3$ & \citet{dorschnerStepsInterstellarSilicate1995a}\\
forst    & forsterite & Mg$_2$SiO$_4$ & \citet{dorschnerStepsInterstellarSilicate1995a}\\
enst     & enstatite & MgSiO$_3$ & \citet{jagerStepsInterstellarSilicate2003a}\\
ferro    & ferrosilite & FeSiO$_3$& \citet{rucksVisibleMidInfraredOptical2022}\\
 fayalite & fayalite & Fe$_2$SiO$_4$ & \citet{jaegerStepsInterstellarSilicate1998}\\
 alumina  & $\theta$ alumina  & Al$_2$O$_3$ & \citet{kurumadaLaboratoryProductionAlumina2005}\\ 
 water & water ice & H$_2$O & \cite{irvineInfraredOpticalProperties1968}\\ % check ref
 spinel  & spinel & MgAl$_2$O$_3$ & \citet{lutzLatticeVibrationSpectra1991}\\% check ref
\enddata
%\tablenotetext{a}{range of grain radii, in $\mu$m}
\end{deluxetable}

Figure~\ref{fig:sixpanel} shows the direct and PSF-subtracted images. 
There are no strong sources in the PSF subtracted images. To be specific, the brightest potential source near the white dwarf has
a surface brightness in the peak pixel of 0.9 MJy~sr$^{-1}$, which corresponds to 6 $\mu$Jy integrated over the PSF area. 
The fluctuations in the 18 and 21 $\mu$m PSF-subtracted images have a gaussian distribution centered at 0, with a width (FWHM) 0.5 MJy~sr$^{-1}$ and 
dispersion $\sigma=0.2$ MJy~sr$^{-1}$. 
The fluctuations appear consistent with random noise. 
A fluctuation greater than 3$\sigma$ would have a brightness of 0.6 MJy~sr$^{-1}$.
Within the central $100\times 100$ pixel region centered on the white dwarf, there are 1100 PSF areas, so the expectation value
is that there could be three random peaks that exceed 3$\sigma$, which is the number seen at each of 18 and 21 $\mu$m.
Furthermore, none of the peaks in the images appear at the same position at more than one wavelength. The planets are expected to be relatively
brighter at the longest wavelengths (but only by modest factors). In terms
of signal-to-noise, planets are expected to be comparable and to appear in multiple wavelengths if real. 

The expected brightness of self-luminous planets or brown dwarfs around GD 362 can be calculated from
models; we use the Sonora models of \citet{marleySonoraBrownDwarf2021} for this purpose.
Assuming an age of 6 Gyr (based on the main sequence lifetime and white dwarf cooling time discussed above),
the mass sensitivity to planets is approximately 25 $M_{\rm Jupiter}$ and effective temperature 500 K. Thus our sensitivity is
rules out bodies larger than late T or Y type brown dwarfs.

\section{Modeling the dust distribution\label{sec:model}}

Dust emission from the GD 362 disk was modeled with {\tt mcfost}, a radiative transfer code
capable of treating the problem of a central illuminating source surrounded by a disk of 
material with potentially high optical depth \citep{pinteMonteCarloRadiative2006}.

The density of dust in the disk is a `fan'-type distribution with a radial power law,
\begin{equation}
\label{eq:density}
n(r,z) = n_1 \left( \frac{r}{r_1}\right)^{-\alpha} e^{-z^2/2H^2},
\end{equation}
{\bfnote between inner and outer radii, $r_1$ and $r_2$, which are also parameters of the model.}
In these equations, $n(r,z)$ is the number density at distance $r$ from the star and $z$ from the midplane; $\alpha$ is the radial power-law exponent; $H$ is the scale height, which flares with increase $r$ as
\begin{equation}
    H(r) = H_1 \left( \frac{r}{r_1}\right)^{\gamma}.
\end{equation}
A flaring exponent $\gamma=1.125$ has been found to fit disks around young stellar 
objects \citep{pinteBenchmarkProblemsContinuum2009}.
The density power law is actually specified via the power-law exponent of the surface density
\begin{equation}
    \Sigma(r) = \int {\rm d}z \, n(r,z) %= \sqrt{2\pi} n_0 r^{-\alpha} H(r) \
    \propto r^{\gamma-\alpha}.
\label{eq:p1}
\end{equation}
Density distributions with $\alpha=1.0$ have successfully fit disks
\citep{pascucci2DContinuumRadiative2004}, 
so the initial guess for the radial surface density power-law exponent 
is  $\alpha\simeq \gamma$.

{\bfnote The emission from each grain of given size is calculated using Mie theory and the 
index of refraction of its constituent material.}
Materials considered for inclusion in the models are those most readily formed
in the hot environments of evolved stars and supernovae, using material of
typical abundances in stellar atmospheres.
To determine the temperature of a particle of given size and composition, the amount
of starlight absorbed and emitted must be balanced, which requires knowledge of
the full imaginary index of refraction (optical constants) from the
ultraviolet through mid-infrared. The materials we include 
in the {\tt mcfost} models are summarized in
Table~\ref{tab:materials}. The size distribution is characterized by a minimum and
maximum grain radius, and a power-law exponent---which was fixed at the value of 3.5,
typical of a collisional distribution of asteroidal debris 
\citep{dohnanyiCollisionalModelAsteroids1969}.

Models are specified by the total mass of dust, the fraction of dust mass in each
constituent, the size distributions for each constituent, the size of the disk,
and the disk geometry exponents. 
{\bfnote 
The fitting of model to data began as an {\it ad hoc} exploration of parameters, but the initial guesses
(based on composition and shape of the GD 29-38 disk) were already reasonably close. 
The model is not very sensitive to some parameters, such as $\gamma$, being somewhat  sensitive to 
$\gamma-\alpha$. 
The model is more sensitive to the scale height at the inner edge, which determines the
optical depth through the disk. We used a coarse grid of $H_1/R_\star=$ 0.5, 1, 5, 18, 30, and 40,
to explore the effect of central optical depth (as discussed in more detail below).
For $H_1/R_\star$=5 and 30, 
we refined  the geometrical parameters using a grid of $r_1$, $r_2$, dust mass,
and composition. Models were compared using their summed squares of deviation from the 
observed spectrum normalized by observational uncertainties.}
The geometrical parameters are summarized
in Table~\ref{tab:params}, and the constituents are summarized in 
Table~\ref{tab:materials}.

\begin{figure}
\begin{center}
\includegraphics[width=.47\textwidth]{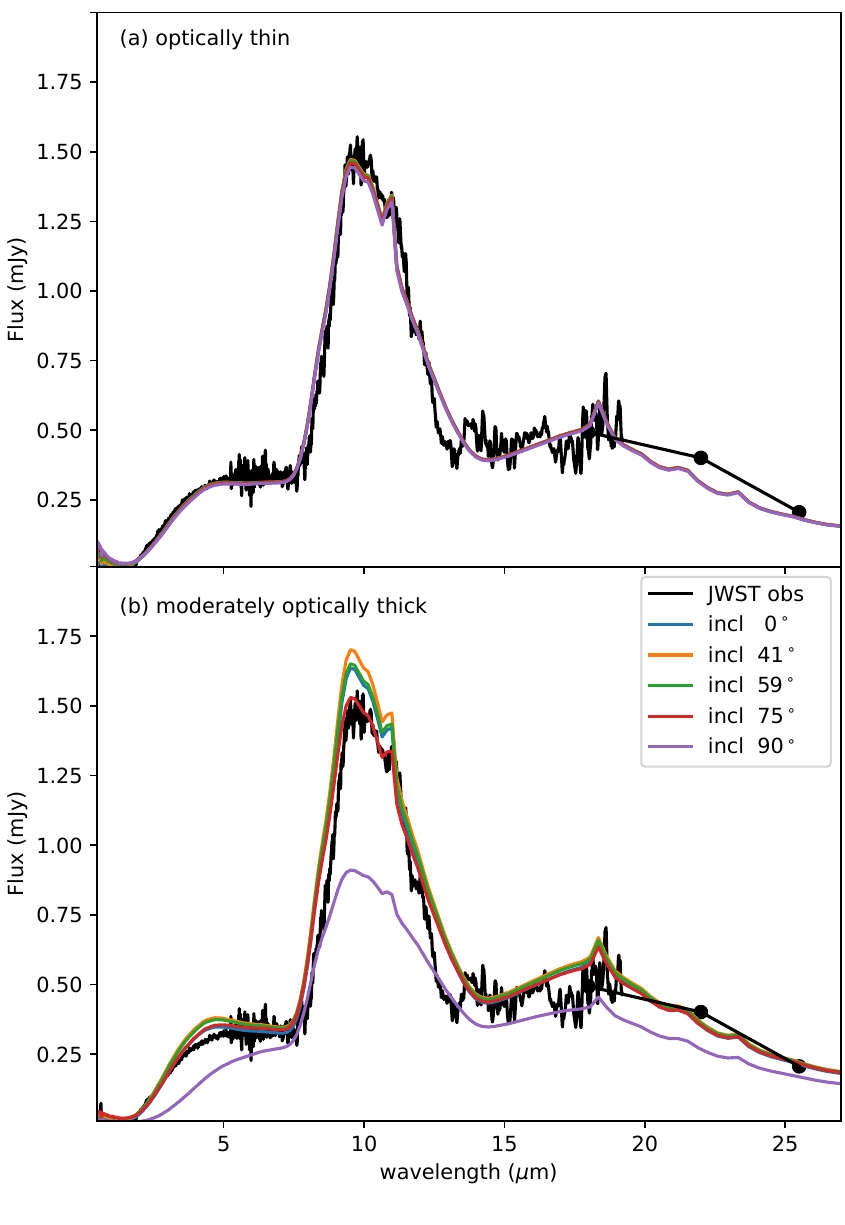}
\end{center}
\caption{Radiative transfer models for the spectrum of dust around GD 362.
(a) The {\it top} panel shows an optically thin model (with a large scale height).
(b) The {\it bottom} panel shows a moderately optically thick model (with a scale height 3.5 stellar radii
at the inner edge).
Grains in both models are composed of `Draine' silicates, amorphous carbon, and forsterite. The dust distribution parameters are in Table~\ref{tab:params}.
 \label{fig:sedfit}}
\end{figure}

To bound the possible materials and disk geometries, we modeled the GD 362 disk using only the smallest
number of minerals and determined which disk properties are well constrained by the observations.
The simplest models with good fits used two commonly found minerals in astrophysical models---specifically, amorphous silicates and carbon.
Those same minerals were found to accurately describe the spectral energy distribution of the brightest white dwarf
debris disk, G29-38 \citep{reachDustCloudWhite2009}.
Figure~\ref{fig:sedfit} shows two models that achieve the best fit,
using as few minerals as possible.
Table~\ref{tab:params} shows the fit parameters.
All models had a surface density radial exponent $p_1=-0.05$ (eq.~\ref{eq:p1}).
The maximum particle size was 2 $\mu$m for good fits; 
increasing beyond this size yielded poor fits as the silicate feature became too weak. 
{\bfnote The minimum size had to be smaller than 1 $\mu$m for a good fit. This makes physical sense, because
the optical parameter $2\pi a/\lambda$ must be smaller than unity to show to show a silicate feature. 
For the prominent feature from GD 362, we found $a_{\rm min}=0.03$ $\mu$m is a good fit, but we did not
explore the precise value.}
The best-fitting
models included interstellar silicates
\citep{draineOpticalPropertiesInterstellar1984},
amorphous carbon \citep{rouleauShapeClusteringEffects1991},
and (crystalline) forsterite \citep{dorschnerStepsInterstellarSilicate1995a},
for which Table~\ref{tab:params} lists the fractional abundance.

\def\extra{
\def\tnm{\tablenotemark}
\begin{deluxetable}{cccccccch}
\tablecolumns{9}
%\tabletypesize{\scriptsize}
\tablecaption{Summary of model parameters\label{tab:params}} 
\tablehead{
\colhead{Model} & \colhead{Mass\tnm{a}} & \colhead{$p_1$} & \colhead{inner-outer\tnm{b}} &  
\colhead{$H_0$\tnm{c}}& \colhead{$\tau_0$\tnm{d}} & \colhead{inclination} &   \colhead{$\chi^2$} &\colhead{comments}
}
\startdata
V2   & 3.6 & -0.05 & 144--1800 & 35 & 0.29 & any         & 1.8 & sil wide 13um\\
V2o  & 1.8 & -0.05 & 162--718  & 35 & 0.29 & any         & 3.2\\
V2o50& 1.8 & -0.05 & 162--718  & 35 & 0.29 & any         & 3.2\\
V2fo & 1.4 & -0.05 & 144--1400 & 31 & 0.28 & any         & 1.9\\
V2fe & 4.6 & -0.05 & 162-2155  & 35 & 0.24 & any        & 5.2   & sil wrong peak\\
V3   & 4.1  & -0.05 & 180-540  & 3.9 & 13.6 & $75^\circ$ & 2.0 & car nar(red)\\
V3o  & 2.2 & -0.05 & 162--718  & 3.5 & 1.16 & $<75^\circ$& 3.4 & too much 18sil\\
V3fo & 1.6 & -0.05 & 215--540  & 4.8 &  & $<80^\circ$    & 2.1 & 60 incl best\\
V6   & 3.0 & -0.05 & 99--1400  & 20 & 0.29 & any         & 2.3 & sil wide(red)\\
V6c  & 3.0 & -0.05 & 90-1400   & 18 & 0.34 & any         & 2.1\\
V6cc & 3.4 & -0.05 & 90-1700   & 18 & 0.26 & any         & 7.0 & sil nar(red)\\
V6p  & 2.7 & -0.05 & 126--1800 & 26 & 0.26 & any         & 9.6 & sil nar(red)\\
v7c  & 2.7 & -0.05 & 90--810   &1.8 & 1.41 & $<80^\circ$ & 2.0 & sil wide(red)\\
v7cp & 3.4 & -0.05 & 110-1100  &2.2 & 1.18 & $<80^\circ$ & 6.5 & sil narr(red)\\
\enddata
\tablenotetext{a}{Mass in units of $10^{19}$ g}
\tablenotetext{b}{inner--outer radii of disk, in units of $R_\star$}
\tablenotetext{c}{scale height of the disk at the inner edge, in units of $R_\star$}
\tablenotetext{d}{optical depth from the star through the disk midplane, at 0.47 $\mu$m}
\end{deluxetable}

\def\tnm{\tablenotemark}
\begin{deluxetable}{ccccc}
\tablecolumns{5}
%\tabletypesize{\scriptsize}
\tablecaption{Summary of model constituents\label{tab:constituents}} 
\tablehead{
\colhead{Model} & \colhead{Constituent} & \colhead{Fraction} & \colhead{small--large\tnm{a}} & \colhead{Topology}
}
\startdata
V2    & DrSil   & 58\% & 0.03--2 & separate\\
      & AmCar   & 39\% & 0.03--1 & separate\\
      & crsil   & 3\%  & 0.03--1 & separate\\
V2o   & olmg50  & 40\% & 0.5-3& separate\\
      & pyrmg50 & 40\% & 0.03--1& separate\\
      & AmCar2  & 18\% & 0.03--1& separate\\
      & forst   & 2\%  & 0.03--1& separate\\
V2fo  & DrSil   & 67\% & 0.03--2 & separate\\
      & AmCar   & 30\% & 0.03--1 & separate\\
      & forst   & 2\%  & 0.03--1 & separate\\
V2fe   & DrSil   & 23\% & 0.03--2& separate\\
      & AmCar   & 43\% & 0.02--1& separate\\
      & ferro   & 54\% & 0.2--2& separate\\
V3    & DrSil   & 74\% & 0.03--2& separate\\
      & AmCar   & 21\% & 0.03--1& separate\\
      & crsil   &  4\% & 0.03--1& separate\\
V3o   & olmg100 & 69\% & 0.5--3  & separate\\
      & AmCar   & 28\% & 0.03--1 & separate\\
      & crsil   &  3\% & 0.03--1 & separate\\
V3fo  & DrSil   & 78\% & 0.03--2 & separate\\
      & AmCar   & 21\% & 0.03--1 & separate\\
      & forst   & 2\%  & 0.03--1 & separate\\
V6    & DrSil   & 62\% & 0.03--2 & mixed\\
      & AmCar   & 34\% & $''$    & mixed\\
      & forst   &  4\% & $''$    & mixed\\
V6c   & DrSil   & 60\% & 0.03--2 & coat\\
      & AmCar   & 40\% & 0.03--2 & coat\\
V6p   & pyrmg50 & 50\% & 0.03--2 & mixed\\
      & AmCar   & 50\% & 0.03--2 & mixed\\
V6p   & pyrmg50 & 50\% & 0.03--2 & mixed\\
      & AmCar   & 50\% & 0.03--2 & mixed\\
\enddata
\tablenotetext{a}{range of grain radii, in $\mu$m}
\end{deluxetable}
}

\subsection{Scale height and optical depth}
{\bfnote A critical parameter for the disk shape, and for calculating radiative transfer of heating radiation, 
is the scale height. A low scale height means a higher optical depth from the star through the midplane.
A large scale height means a lower optical depth. To match the observed brightness of the GD 362 disk, the 
the optically depth to visible light from the star
through the midplane becomes greater than unity for a scale height $H_1 \le 10 R_\star$. 
We explored models spanning this divide.}
\def\extra{The scale height of a disk near GD 362 that remains optically thin is 30 $R_\star$ for a disk that extends inward to $100~R_\star$.} 
We started with a scale height 31 $R_\star$ for the optically thin models, then decreasing to 4.8 $R_\star$, where the disks is moderately optically thick in order to generate the
observed brightness. 
The {\tt mcfost} model can handle high optical depths, which were also calculated.
Models with a low scale-height, $H_1=0.5 R_\star$
at the inner edge, were so optically thick that they were unable to explain the observed emission,
yielding relatively smooth spectra lacking the 10 $\mu$m silicate feature that dominates the observed spectrum.

\def\tnm{\tablenotemark}
\begin{deluxetable}{ccc}
\tablecaption{Summary of model parameters\label{tab:params}} 
\tablehead{
\colhead{} & \multicolumn{2}{c}{Model}\\ \cline{2-3}
\colhead{Parameter} & \colhead{Optically thin} & \colhead{Moderately thick}
}
\startdata
Dust mass & $2.7\times 10^{19}$ g & $1.6\times 10^{19}$ g\\
Inner radius ($r_1$) & 140 $R_\star$ & 215  $R_\star$\\
Outer radius ($r_2$) & 1400 $R_\star$ & 540  $R_\star$\\
Scale height ($H_1$) & 31 $R_\star$ & 4.8  $R_\star$\\
Scale height exponent ($\gamma$) & 1.125 & 1.125\\
Surface density exponent ($\gamma-\alpha$) & -0.05 & -0.05 \\
Midplane optical depth & 0.28 & 6.5\\
Inclination  & any & $<80^\circ$\\
DrSil\tnm{a} & 67\% & 78\% \\
AmCa\tnm{a}   & 30\% & 21\% \\
forst\tnm{a}        & 2\%  & 2\% \\
grain size & 0.03--2 $\mu$m  & 0.03--2 $\mu$m\\
\enddata
\tablenotetext{a}{Fraction of mass in these materials; see Table~\ref{tab:compositions} for definitions. }
\end{deluxetable}

Higher optical depth models, with scale height less than the stellar radius, do not fit the observed spectrum.
For one thing, the optically thick models do not produce a 9--11 $\mu$m silicate feature.
Models with scale height less than 0.5 $R_\star$ cannot
match the observed brightness at any mid-infrared wavelength: such
disks cannot absorb enough starlight to explain the high luminosity of the disk.
(To see this analytically, a flat disk can absorb a maximum 
fraction $4/(3\pi r/R_\star)$ of the star's light, which requires 
an inner radius of the disk $18 R_\star$ well within the Roche and sublimation limits.) 

The emission from GD 362 is attributed to carbon grains in this model, because carbon is an abundant material
and the grains are hotter than silicates, matching the observed spectrum very well with the same
spatial distribution for both the carbon and silicate grains.
However, it is also possible that some of the near-infrared emission is from the inner edge of a 
high-optical-depth disk (embedded in the disk that explains the bulk of the mid-infrared emission).
Temperatures become so low that the disk does not emit significantly in the mid-infrared. 
The small grains responsible for the silicate feature must contribute at least half of the observed flux. 
The inner edge of a flat, optically thick disk such as described by \citet{juraInfraredEmissionDusty2007} could explain some of the emission, in particular in the near-infrared. 
There is a tradeoff between a flat disk and carbonaceous grains.
In what follows, we consider the near-infrared as arising from carbonaceous material thati s cospatial
with the silicates (for which the emission must arise from moderate-to-low optical depth disk).

\subsection{Inclination}
The optically thin model is for a specified inner scale height but can likely be fit equally well with a spherical shell. The optical depth through the midplane is 0.28 if the scale height is {\bfnote 31} $R_\star$. Any inclination of the disk 
provides approximately the same observed spectrum for the optically thin model. 
The scale height tells us the distribution of inclinations of the
orbits, and for the optically thin model it corresponds to a parent
body (or width of distribution of parent bodies) with $13^\circ$
inclination, typical of Solar System asteroids. 

The moderately optically thick model has a midplane optical depth  of 6.4
(at a wavelength of 1 $\mu$m),
which is high enough that the observed flux depends upon the angle at which the disk is
viewed. Figure~\ref{fig:sedfit}b shows the models for five inclinations,
from face-on ($0^\circ$) to edge-on ($90^\circ$). The edge-on model does
not fit as well, because the midplane is colder. All models in Fig.~\ref{fig:sedfit}b have the same dust mass and spatial distribution.
For the edge-on model, an improved fit is obtained by moving the inner radius further inward
and increasing the dust mass, but the fit to the spectrum is still not 
as good as the lower-inclination models.

{\bfnote 
The star can be occulted by the disk, if it is seen at high inclination.
For the moderately optically thick model, $H_0\simeq 3 R_\star$, the disk 
completely occults the star if viewed at inclination $i>88.6^\circ$. 
For a larger scale height, $H_0=35 R_\star$, the disk partially blocks the star if $i>73^\circ$,
with midplane optical depth of 0.3 at 0.47 $\mu$m and 0.5 at 0.1 $\mu$m.
The white dwarf does not appear to have such high extinction, so the disk is 
unlikely to have inclination higher than $73^\circ$ and certainly not higher than $88.6^\circ$.
Other white dwarfs with dust disks that are more nearly edge on may show transit effects \citep{bhattacharjeeThickDisksWhite2025}.
}

\subsection{Inner and outer radius}
The inner edge radius is largely set by the color temperature of the continuum from 2--7 $\mu$m.
There is an additional, mild
dependence of the inner radius on the optical depth of the disk, because optically thick disks have
cooler midplanes. 
We found the inner radius for all reasonable-fitting models to be 
around 100 $R_\star$, comparable to the Roche limit for tidal
disruption (82 $R_\star$).

At wavelengths shorter than 4 $\mu$m, the model emission is
primarily from amorphous carbon, because the silicates are (in comparison)
relatively transparent and remain cooler at the same distance from the star.
In the ISM, carbonaceous material is characterized by a set of polycyclic aromatic hydrocarbon spectral
features, including strong ones at 3.3 and 6.2 $\mu$m. 
However, 
the photon arrival interval for the dust grains near the white dwarf is high enough (compared
to their cooling times) that the grains achieve a steady-state temperature.
Thus, unlike in the ISM, the grains do not 
experience thermal pulses that allow the mid-infrared hydrocarbon features to be prominent \citep{dwekDetectionCharacterizationCold1997}.
{\bfnote The lack of 3.3 and 6.2 $\mu$m features in the GD 362 spectrum could just mean
that hydrocarbon grains are not present.
In any event, the emission near GD 362 can be characterized using bulk optical
constants (for both silicates and carbon) and an equilibrium temperature.}

The outer radius of the disk is determined by the falloff in brightness at the longest wavelengths of the spectral energy distribution. The JWST observations extend to 25.5 $\mu$m, and the disk emission
is clearly decreasing at wavelengths longer than 18 $\mu$m. Thus the outer radius is reasonably well constrained by the decrease in brightness from 18--25 $\mu$m, with some dependence on the index
of refraction of the minerals across those wavelengths. We find the outer radius of models that
reasonably fit the data to be in the range 540--1400  $R_\star$ (0.03--0.08 au) depending
on the disk optical depth.

Material further from the star than our model disk could harbor mass
that is not detectable.
Dust further than 1 au from the star would have a temperature 60 K, which would make no
contribution to the mid-infrared emission.
Planetary material is anticipated around the main sequence progenitor 
at a range of distances out to $10^4$ au, based on  analogy to 
the Solar System. 
However, the material closer than 1.8 au was  engulfed in the
progenitor during its asymptotic giant phase, based on the MESA Isochrones and StellarTracks (MIST) models \citep{choiMESAISOCHRONESStelLAR2016}
for the GD 362 progenitor mass (Table~\ref{tab:gd362}).
If the white dwarf comprises half of the progenitor star
from that phase, then orbits just beyond engulfment would
have expanded to 3.6 au once the outer envelope was lost. 
Thus it is anticipated that there
is not significant relic planetary bodies closer than
3.6 au from the star.
An analog to the zodiacal dust cloud in the Solar System could exist,
and it would fill in the inner part of the GD 362 planetary system as
material is produced by collisions then spirals inward under Poynting-Robertson drag.
However, to reprocess the luminosity of GD 362's disk, a zodiacal cloud analog would need to be 5 orders of magnitude more dense than the zodiacal cloud, which has optical depth only of order $10^{-7}$  \citep{reachZodiacalEmissionDust1988,kelsallCOBEDiffuseInfrared1998}.

In the asteroidal disruption model \cite{juraPollutionSingleWhite2008}, the GD 362 disk material 
was transported from further out, via gravitational perturbations of
planetesimals by a giant planet. 
A planetary body that arrives within the star's Roche limit
(82 $R_\star$ for For GD 362) is tidally disrupted. 
%If the body had an 
%apastron of 3.6 au and periastron of the Roche radius, then it would
%have a high orbital eccentricity of 0.997.
We expect the debris from the disrupted parent body to be initially spread from
82 $R_\star$ to at least 3.6 au, but the orbits of the collisional
products  become rapidly circularized \citep{verasFormationPlanetaryDebris2015}
leading to orbits just outside the Roche radius.
For comparison, our  disk model, based only on spectral fitting,
has an inner radius of 140 $R_\star$ and outer radius 1400 $R_\star$ (0.08 au);
this likely captures the bulk of the debris within the 
context of this model.

\subsection{Disk mass\label{sec:mass}}

The key to determining the total mass is including the largest bodies.
Because of the steep dependence of a collisional size distribution on particle size, 
the total surface area (which determines the observed brightness) depends on the minimum grain radius, while the total mass depends more on the largest grain radius. 
For a size distribution with number density depending on radius ($a$) as $a^{-3.5}$
\citet{dohnanyiCollisionalModelAsteroids1969}, the 
mass scales with $a_{\rm max}^{0.5}$. 
For a maximum grain radius of 2 $\mu$m, we find that the mass of 
small grains in models that can match the JWST
observations is  $M_{\rm d19}\simeq 2\times 10^{19}$ g.
If  this mass were compressed into a solid body with
with density 3 g~cm$^{-3}$, its effective radius would be 12 km, like 
a small asteroid.

Including bodies up to 1 km and using a Dohnanyi size distribution
can approximately reproduce 
the overall spectral energy distribution (with inner and
outer radius of 72--540 $R_\star$) with a total mass of $1\times 10^{21}$ g,
albeit with a silicate feature too weak by a factor of 2.5.
Experiments using size distributions with slope $p=3.4$ and 3.6 yielded
comparable results of too-weak silicate features,
indicating that the observed emission is primarily from small dust.

This likely means that the size distribution changes slope (becomes steeper)
at sizes above 2 $\mu$m and below 1 km. Such a transition is
expected because solids change from strength-dominated to rubble piles,
and interplanetary particles larger than 2 $\mu$m have significant porosity \citep{bradley126InterplanetaryDust2007}, which 
affects their collisional strength \citep{loveTargetPorosityEffects1993}.
The total mass is sensitive to the power-law index, 
and hyper-velocity impact experiments indicate $3.5<p<3.7$ \citep{takasawaSilicateDustSize2011}.

In principle, we can extrapolate the size distribution upward to where
there is a single body to get the total mass.
These estimate of the parent body size are only for illustration because
of the wide range of size extrapolating from dust to planetary-sized bodies.
For a size distribution with size-frequency index $p=3.7$, 
the effective radius of the parent body is 190 km, equivalent to a large asteroid.
This calculation is very sensitive to the poorly known value of $p$.
For $p=3.5$, the effective radius would be 1400 km (the size of a dwarf planet), 
but we already ruled out $p=3.5$ continuing above 1 km size based on the 
spectral shape. 

In summary the mass of solids around GD 362 must be at least the mass of small grains,
$2\times 10^{19}$ g, and for a plausible size distribution, the total mass is likely of order $10^{22}$ g.

For comparison, a large amount of planetary material is inferred to have been
accreted by the white dwarf, in order to account for the metal abundance in
its atmosphere. 
Analysis of the material accreted onto polluted white dwarf
atmospheres suggests they accreted an amount of material equivalent to
parent bodies with masses of order $10^{23}$ g and 
equivalent radii of order 200 km
\citep{farihiCircumstellarDebrisPollution2016,harrisonEvidencePostnebulaVolatilisation2021}. 

The disruption of bodies larger than 30 km can explain the polluted
atmospheres of DB white dwarfs like GD 362 \citep{wyattStochasticAccretionPlanetesimals2014}.
The disk mass (including larger solids) can be explained by the
collisional cascade models of \citet{kenyonNumericalSimulationsCollisional2017},
where a steady-state disk of the observed mass if the parent bodies of
100 km radius disrupted. 

\subsection{Comparison of GD 362 to G29-38 disk shape}

\begin{figure}
\begin{center}
\includegraphics[width=.499\textwidth]{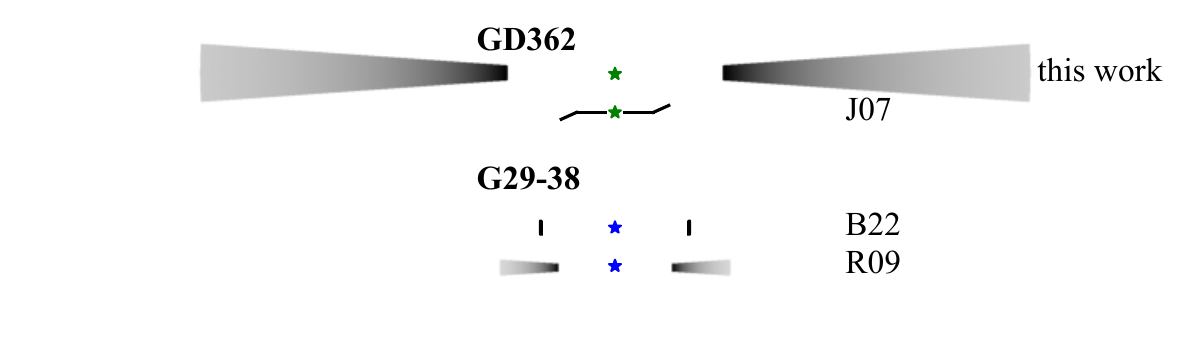}
\end{center}
\caption{Comparison of the density distributions for models fit to G29-38 (lower two models) and GD 362 (upper two models).
The scales and references are listed in Table~\ref{tab:comparison}.
For G29-38, the model from B22 is a ring (vertically thick but radially narrow), at a radius within the range of 
the power-law torus of R09. For GD 362, the model from J07 is a flat, opaque disk that warps beyond 50 $R_\star$.
\label{fig:comparison}}
\end{figure}

The white dwarf G29-38 is similar to GD 362, with photospheric metal absorption lines as well as thermal emission
from a disk of solid material. Table~\ref{tab:comparison} summarizes size scales inferred from modeling dust
around the two stars, and Figure~\ref{fig:comparison} shows the models graphically. For G29-38, the two models
shown are in reasonable agreement, within the bounds of their assumptions. The larger dust mass in R09 compared to B22
is at least partially due to the larger range of distances from the star in the former model. 
For GD 362, the J07 model is for a flat disk (scale height much less than the stellar radius), which 
was then warped so that the outer portions receive starlight to produce the mid-infrared emission. The model
in the present paper is quite different, with the material further from the star and

\def\tnm{\tablenotemark}
\begin{deluxetable}{cccccc}
\tablecaption{Comparison between G29-38 and GD 362 models\label{tab:comparison}} 
\tablehead{
\colhead{} & \multicolumn{2}{c}{G29-38} &&  \multicolumn{2}{c}{GD 362} \\ \cline{2-3}\cline{5-6}
Parameter & R09\tnm{a} & B22\tnm{a} && J07\tnm{a} & this work}
\startdata
$r_1$ ($R_\star$) & 74  & 96                 && 12 & 140  \\
$r_2$ ($R_\star$) & 150 & 97                 && 70 & 540  \\
$H_1$ ($R_\star$) & $>2$ & 8                 && $<1$& 4.8 \\
$M_{\rm dust}$ ($10^{18}{\rm g}$) & 20 & 4.5 && $>1$ & 20  \\
\enddata
\tablenotetext{a}{References: R09 \citep{reachDustCloudWhite2009}, B22 \citep{balleringGeometryG2938White2022},
J07 \citep{juraInfraredEmissionDusty2007}}

\end{deluxetable}

\section{Modeling the Dust Mineralogy\label{sec:mineralogy}}

To determine the detailed mineralogy, we now focus on the details that are
evident in the new JWST spectrum of GD 362.
We used laboratory measured emissivity spectra of over 100 candidate materials typically found in astrophysical dust, comet sample returns, and meteorites.
The emissivities are combined linearly, with an adjustable effective temperature of the grains, in order to best match the observed spectrum. 
The absorption cross-sections were calculated for a range of dust particle sizes, 
and adjustments are made to the effective particle temperature and emissivity from the fiducial 1 $\mu$m particle size of the laboratory measurements {\bfnote using Beer's Law, as
described in  detail in two prior papers using the same technique
\citet{lisseSpitzerSpectralObservations2006,lisseAbundantCircumstellarSilica2009}. 
We experimented with power law size distributions
(with and without a $<1$ $\mu$m particle-size rolloff due to radiative blowout effects) before settling on the best fit $dn/da \propto a^{-3.7}$ over the size range
$0.1 < a < 1000$ $\mu$m.
}
Similar analyses have been performed and published for comets \citep{lisseComparisonCompositionTempel2007,sitkoInfraredSpectroscopyComet2011},
debris disks \citep{lisseCircumstellarDustCreated2008,lisseSpitzerEvidenceLateheavy2012,lisseHD145263Spectral2020},
and most applicably, for the polluted white dwarf system G29-38 \citep{reachDustCloudWhite2009}.
We find an overall continuum temperature of 950 K converts the observed JWST fluxes into well-flattened emissivities to be studied for relative contributions of the individual mineral components. Note that at 950 K, the temperature is high enough to affect, via melting and annealing, metal sulfides and magnesium-rich pyroxenes.

\begin{figure*}
\begin{center}
\includegraphics[width=\textwidth]{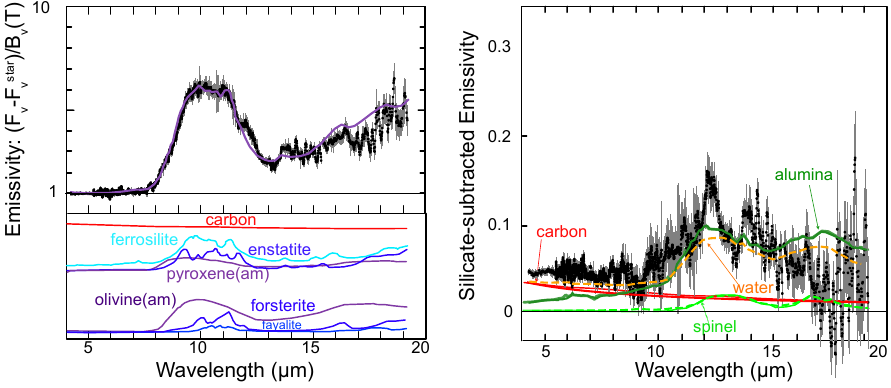}
\end{center}
\caption{Mineralogical decomposition of GD 362's mid-infrared spectrum. 
{\it (left)} The upper-left panel shows the observed spectrum (black with grey error bars) together with the model
fit (purple). The lower-left  panel shows the contributions from each silicate mineral in the model.
{\it (right)} The right panel shows the observed emissivity spectrum after subtracting the modeled silicates (black with grey error bars). Contributions from non-silicate species are shown, with dashed lines for  components that were not included in the
final model because their contributions did not significantly improve the fit (as discussed in the text).
\label{fig:compositions}}
\end{figure*}

Figure~\ref{fig:compositions} summarizes the mineralogical decomposition, and Table~\ref{tab:compositions} lists the minerals in order of their significance to the fit. 
The fit is not perfect ($\chi^2=1.55$) with some systematic residuals in the 11--13 $\mu$m range that are evident after removing the dominant silicates.
The values of $\Delta\chi^2$ in Table~\ref{tab:compositions} are the change in the goodness of fit if the mineral is removed.
The decision of whether a mineral contributes significantly to the fit is based on
an F-test comparing the $\chi^2$ with and without the mineral. For $10^4$ degrees of freedom (the number of spectral points), the improvement from including the mineral is 95\% likely to have improved the fit (as opposed to being a chance match to measurement errors) if $\Delta\chi^2>0.05$. This means the top 8 minerals in Table~\ref{tab:compositions} significantly improved the fit. The last two are included
for completeness, showing that they were not close to criterion for significance.
We find the emission can be explained as arising from a combination of ferromagnesian olivine and pyroxene, mixed with amorphous carbon soot. Many other species were attempted but had improved the fit
less than the species listed here. 
There could also be other solid species that have no noticeable mid-infrared features. For example, many solid elements such as iron or nickel could replace the amorphous carbon (as discussed in the next section).

Silicate dust dominates, at 83\% of the total material; about 1/3 of this is in an amorphous, glassy state. This shows how distinct the material around GD 362 is from
the interstellar medium, where nearly all silicates are amorphous \citep{kemperAbsenceCrystallineSilicates2004}.
The calculated olivine to pyroxene ratio is $\sim$0.9 from the derived individual species surface area tabulated in Table~\ref{tab:compositions}, indicating a relatively primitive parent body. No obvious evidence for phyllosilicates was found, suggesting little to no aqueous alteration occurred on the parent body. No evidence for silica was found, suggesting no highly energetic impact processing occurred when producing the dust from the parent body. 
Of the silicates, the olivines are 55\% amorphous, while the pyroxenes are only 23\% amorphous; this could be
attributed to the lower melting and annealing temperatures of pyroxenes.

\begin{deluxetable}{cccc|cccc}
\tablecolumns{8}
\tablecaption{Dust species from spectral modeling\label{tab:compositions}} 
\tablehead{
\colhead{Mineral} & \colhead{Formula} & \colhead{$\Delta\chi^2$} & \colhead{Area fraction} & 
  \colhead{Density} & \colhead{Mass fraction} & \colhead{Molar mass} & \colhead{Molar fraction}\\
(1)       & (2)                          & (3)  & (4)  & (5) & (6)    & (7) & (8)  }
\startdata
ferrosilite       & FeSiO$_3$                     & 26.3 & 18\% & 4.0 & 28.3\% & 264 & 11.7\% \\ %FerroS
olivine (am.)     & MgFeSiO$_4$                   & 12.0 & 17\% & 3.6 & 24.1\% & 172 & 15.3\% \\
carbon (am.)      & C                             &  7.8 &  5\% & 2.5 &  4.9\% &  12 & 44.8\% \\
pyroxene (am.)    & MgFeSiO$_3$                   &  3.2 &  7\% & 3.5 &  9.6\% & 232 &  4.5\% \\
enstatite         & MgSiO$_3$                     &  2.6 &  6\% & 3.2 &  7.6\% & 200 &  4.1\%\\ % clinenst
$\theta$ alumina  & Al$_2$O$_3$                   &  1.6 &  5\% & 4.0 &  7.9\% & 150 &  5.7\%\\
forsterite        & Mg$_2$SiO$_4$                 &  1.5 & 10\% & 3.2 & 12.6\% & 140 &  9.8\%\\ %ForstK
fayalite          & Fe$_2$SiO$_4$                 &  0.08&  3\% & 3.2 &  5.1\% & 204 &  4.0\%\\ %FayaJ
water ice\tnm{a}         & H$_2$O                        &  0.02&  1\% & 1.0 &  0.4\% &  18 &  2.3\%\\ % possibly remove from this table because not significant
spinel\tnm{a}            & MgAl$_2$O$_4$                 &  0.01&  1\% & 3.6 &  1.4\% & 142 & 1.1\%\\ % possibly remove from this table because not significant
\enddata
\tablenotetext{}{Description of columns: 
(1) Name of the species or mineral. 
(2) Chemical formula, for stoichiometry.
(3) Change in goodness-of-fit if this species is removed; the total $\chi^2$ is 1.55.
(4) Fractional contribution to the total grain surface area.
(5) Bulk density of the dust species, in g~cm$^{-3}$.
(6) Fractional contribution to total observed dust mass.
(7) Molar mass in amu.
(8) Fractional molar abundance of each species.}
\tablenotetext{a}{Water ice and spinel are listed here for completeness, but their $\Delta\chi^2$ is not statistically significant as described in the text. These 
materials were therefore not included in the final abundances.}
\end{deluxetable}

{\bfnote
The only non-silicate minerals in the fit are carbon and alumina (Al$_2$O$_3$).
Alumina has three common, stable allotropes ($\alpha$, $\delta$, and $\theta$),
of which we only found evidence for $\theta$ in our spectra. 
Alumina  is not as commonly reported as silicates or carbon
in presolar material, but it is both a predicted and observed mineral.
Alumina is directly observed in presolar grains in meteorites; alumina
and spinel are the top two known compositions of presolar (non-silicate) oxide grains
\citep{nittlerPresolarOxideGrains2025}.
Indeed, alumina is predicted to be the first abundant solid to form in O-rich stellar atmospheres (where silicates are formed) due to its
high condensation temperature. Alumina grains may form the seeds for formation of
silicates. In the total mass loss from O-rich evolved stars, silicates are more abundant, due to the higher abundance of Si relative to Al. Furthermore, silicates
dominate the infrared emission because of the intrinsic strength of the 9--11 $\mu$m 
feature that overlaps with the signature of alumina \citep{sloanInfraredSpectralClassification1998}. 
The abundance of alumina is not well quantified in presolar material, but the
about of aluminum found in the GD 362 spectral fit is in keeping with 
that seen in the photosphere of GD 362 itself, and comparable to 
the abundance in meteoritic material, as discussed in the following sections. 
}

\subsection{Stoichiometry}

To calculate the elemental abundances in the solids around GD 362, we first determined the molar abundance
of each species, then broke it down into each element. Table~\ref{tab:compositions} contains the
per-species results. We used the fitted fractional surface area of each 
dust species (column 4), together with
its bulk density (column 5) to get the mass fraction (column 6). 
Then with the chemical formula (column 2), we calculate the molar
mass (column 7), and then the relative number of moles of each dust species (column 9).

\begin{deluxetable}{lhcccccc}
\tablecolumns{9}
\tablecaption{Abundances in Material around and in GD 362 \label{tab:stoichiometry}} 
\tablehead{
& \multicolumn{7}{c}{Abundance\tnm{a}}\\ \cline{2-8}
\colhead{Source} & 
%\colhead{H}& 
&
\colhead{C} & \colhead{O} & \colhead{Mg} & \colhead{Al} & \colhead{Si} & \colhead{Fe} 
}
\startdata
%                              H         C      O      Mg     Al   Si    Fe  
dust disk\tnm{b}           & $<.004$ & 0.28 & 2.23 & 0.60 & 0.16 & 1 & 1.42 \\
$\div$meteoritic\tnm{c}    & $<.03$  & 0.73 & 0.53 & 0.68 & 2.07 & 1 & 0.82 \\
\\
photosphere\tnm{d}         & 50,000  & 0.14 & $<5$ & 0.72 & 0.21 & 1 & 1.55 \\
$\div$meteoritic\tnm{c}    & 290,000 & 0.09 &$<0.4$& 0.39 & 2.68 & 1 & 2.02  \\
\enddata
\tablenotetext{a}{Abundance by mass, relative to Si, for elements in minerals that were statistically significant in fitting the JWST disk spectrum.}
\tablenotetext{b}{this work}
\tablenotetext{c}{\citet{loddersRelativeAtomicSolar2021}}
\tablenotetext{d}{ratio of accretion rates \citet{zuckermanChemicalCompositionExtrasolar2007,xuTWOBEYONDPRIMITIVEEXTRASOLAR2013}}
\end{deluxetable}

The elemental abundances we calculated from the relative molar abundance, weighted by the chemical formula of each
species and mass of each element.
Table~\ref{tab:stoichiometry} lists the atomic abundances (per unit mass) for the elements contained in the dust-bearing species
from the mid-infrared spectral fitting. The most abundant element, both by mass and number of atoms, is oxygen; this is not surprising given the prevalence of silicates.

The second row of Table~\ref{tab:stoichiometry} compares the abundances to 
those measured in CI chondrites \citep{loddersRelativeAtomicSolar2021}, which derive from `carbonaceous' asteroids \citep[which in 
turn may have derived from comets, i.e. formed in the outer solar system, per][]{nesvornyCometaryOriginZodiacal2010}. The abundance pattern is roughly similar between
GD 362 and chondritic material, except the oxygen is low and the aluminum is high (both by factors of approximately 2). A low abundance of an element relative to chondritic could mean that the composition of the material around GD 362 differs, or the element is present in other forms. For oxygen, other carriers
that are not active in the mid-infrared could be present, or the oxygen could have been lost from the solid phase 
as vapor during collisional processing. 

An element notably missing from the dust inventory, given its abundance
in meteorites, is sulfur. Sulfur is volatile and may also have been lost as vapor.
{\bfnote Meteoritic sulfur is troilite (FeS), pyrite (FeS$_2$) and elemental, and {\it en route} to 
Earth was exposed to sunlight of comparable intensity to that in the GD 362 disk.
Magnseium/iron sulfides have prominent far-infrared emission features at 33--38 $\mu$m, outside of the JWST spectra range.}

%The high abundance of aluminum depends upon the identification of alumina in the spectral fits. Because it was only 
%a trace component of the mid-infrared spectral fit, we do not put high credence in this apparently anomalous abundance.

The relative abundance of Mg and Fe depends upon their abundances within the silicate
minerals. While the crystalline materials are distinct in the mid-infrared, the amorphous
forms have spectral shapes that depend only weakly upon the relative amounts of Mg 
and Fe.
For the amorphous olivine and pyroxene material, we assumed equal fractions
of Mg and Fe, which yielded a mass abundance ratio Mg/Fe=0.42. 
If we change the amorphous silicate material to be fully Mg-bearing, the Mg/Fe ratio changes to 1.12. If we change the amorphous silicate material to be fully Fe-bearing,
the Mg/Fe ratio changes to 0.20. Neither of these extremes is likely; using a Mg fraction of 50\%$\pm$25\% in the amorphous silicates, the range of Mg/Fe ratios in
the total dust mixture is 0.28--0.65. For comparison, the 
chondritic ratio is 0.51, which is well within the range for GD 362. Indeed, an 
exactly chondritic ratio of Mg/Fe is obtained if the amorphous silicates have
63\% Mg and 37\% Fe.

As mentioned above, the amorphous carbon in the spectral fit could be substituted with other, largely spectrally neutral, 
materials. Based on cosmic abundances, the common elements not accounted for in the dust include 
light, highly volatile ones like H and N (and of course, the noble gases, which do not readily form compounds). 
The other abundant elements include S and Ni. 
Furthermore, 18\% of the highly-abundant Fe is not  accounted, and that fraction increases to 54\% if the amorphous silicates
are purely magnesium-rich.

\subsection{Comparison of GD 362 dust to photospheric metals}
The elemental abundances inferred from the gas in GD 362's photosphere are
listed in the third row of Table~\ref{tab:stoichiometry}, using the accretion rate of each element \citet{zuckermanChemicalCompositionExtrasolar2007,xuTWOBEYONDPRIMITIVEEXTRASOLAR2013}. 
(O has only upper limits from these observations.) 
%A white dwarf atmosphere
%is expected to be composed almost entirely of H and He, because heavier elements rapidly settle
%downward form the high-gravity conditions near the surface of the star.  #Muk: GD362 has a helium-dominated atmosphere, so the gravitational settling times would be longer than for DA white dwarfs.
Photospheric metal abundances for polluted white dwarfs are typically similar to chondrites \citep{farihiCircumstellarDebrisPollution2016,swanPlanetesimalsDZStars2023,xuChemistryExtrasolarMaterials2024}.
For GD 362, the photospheric accretion abundances
are generally similar to chondritic material and the GD 362 dust, except that C has far lower accretion rate, relative to Si, than the chondritic abundance ratio of those elements. A similar abundance pattern was found for another heavily polluted white dwarf, WD 1145+017 \cite{lebourdaisRevisitingChemicalComposition2024}.

We used CI chondrites as the template for solar system solids, which are the 
most carbon-rich meteorites.
In fact there is a range of observed compositions in planetary materials, with carbon being
much lower. 
Using the abundances of many elements in the photosphere, including  trace elements that could be observed
in the gas phase but would not be detectable in dust, \citet{xuTWOBEYONDPRIMITIVEEXTRASOLAR2013} found the abundance pattern 
for GD 362 best matched `mesosiderite' meteorites. Those rare meteorites lack olivine,
while there are strong olivine features in the JWST dust spectrum.
The solids seen around GD 362 have abundances that do not
match any particular type of meteorite well, which is not surprising given that we
are studying the amalgam of surviving material from a disruption event around
the white dwarf, rather than individual rocks that can arise from a specific portion of a parent body.
A geochemical analysis shows that some of the white dwarf atmospheric abundance patterns
match no known meteorites \citep{xuChemistryExtrasolarMaterials2024}.
An individual meteorite may derive from only the core, mantle, or other part  of the
parent body, while the collisional comminuted dust  contains a mixture of all
parts of the parent body---and potentially even multiple parent bodies.

\subsection{Comparison of GD 362 to G29-38 mineralogy}
The elemental abundances in the dust around GD 362 can be compared to those found for G29-38 using its Spitzer spectrum \citep{reachDustCloudWhite2009}. 
Compared to GD 362, G29-38 dust has the same abundance of O and Mg, relative to Si. This suggests that silicates of comparable net stoichiometry exist in the planetary material around both stars, most likely due to the ubiquity of such material formed in the atmospheres of red giant stars and seeding the material
from which GD 362 and its planetary system formed.
No Al-bearing species was found in the G29-38 dust mineralogy, so we cannot
compare Al abundances quantitatively, but note that it is likely lower in dust around G29-38. A Ca-bearing mineral (diopside) and S-bearing mineral (noningerite) were found around G29-38 though no such minerals were found in the
fit for GD 362, so we do not compare those elements.
Fe may be more abundant in G29-38 dust than GD 362 dust (by a factor 1.8). 
Relative to CI chondrites, Fe is 18\% under-abundant for GD 362 and 49\% over-abundant for G29-38. 
The C abundances inferred from our dust models were different for the two stars: 
relative to chondrites, C is 72\% under-abundant for GD 362 and 65\% over-abundant for G29-38.

The similarities and differences in elemental abundances may indicate differences in the parent
bodies of the GD 362 and G29-38 dust. 
GD 362 may be more consistent with a C-type asteroid.
The lack of water or phyllosilicates in the GD 362 material also support a dry,
C-type parent body.
In contrast, the higher C abundance around G29-38 may reflect a composition more consistent with cometary material.

\section{Discussion\label{sec:discussion}}

\subsection{Water in the GD 362 Disk?}

There is no strong evidence for any H$_2$O gas or ice in the spectrum of GD 362, although H$_2$O has prominent
features within the observed wavelength range.
When included in the fits, the H$_2$O ice mass is $4\times 10^{16}$ g.
Given the lack of significant improvement in the fit when H$_2$O is included, we instead interpret this as an upper limit of
$8\times 10^{16}$ g at a 95\% confidence level.
{\bfnote Lack of water ice is expected, as it would be desorbed from grains at their high temperature (950 K) 
in the GD 362 disk.}
Note that this upper limit is in the context of the model for
an optically thin dust cloud, with all species sharing the spatial distribution of the dominant silicate species. 
In terms of elemental abundances, much more O is in the silicates than water, but water is the only H-bearing species in the model. The hydrogen abundance is discussed in the next subsection.

The material around GD 362 could have been dry, which is 
consistent with the lack of any detected aqueous alteration products in the silicate mineralogy.
However, any water in the system would have a very short lifetime if exposed to the radiation field of GD 362 at the
distances of the disk from the star.

The lifetime of H$_2$O in the GD 362 disk is estimated as follows.
For reference, at 1 au from the Sun, the photodissociation time is 0.96 days when the Sun is quiet and 0.53 days when it is active \citep{huebnerPhotoionizationPhotodissociationRates2015}.
so the photodissociation time around GD 362 can be estimated by scaling from the solar values, using the relative  brightnesses of the stars. 
%For a crude estimate, the ratio of blackbodies at the effective temperatures of GD 362 and the Sun is 1200 (in favor of GD 362),
%and scaling by the total luminosity of the much smaller white dwarf results in GD 362 being about 1.9 times brighter than the Sun at 1500 \AA. 
%We can improve this estimate using the actual UV flux of the star.
The most important wavelengths for H$_2$O dissociation correspond to the GALEX FUV channel, which spans 1340-1809 \AA.
For the Sun, the SOLAR-ISS \citep{meftahSOLARISSNewReference2018} shows the flux at
these wavelengths is 
%1e-3 W/m2/nm which corresponds to 
$7.7\times 10^9$ Jy
%, or magnitude -15.8.
at 1 au distance.
%G29-38:
%From the GALEX MAST catalog the magnitude is 15.14 in the FUV. That means 0.0032 Jy seen from Earth.
%Moving it out to 14 pc distance, it would have a magnitude 16.5. So the Sun is about 10**(.4*(16.5-15.14))=3.5 %fainter than the WD for a body 1 au from it.
%That means the photodissociation time at the 0.01 au from the WD would be 0.96 days * .01**2 / 3.5 = 2.5 sec.
For GD 362, from the GALEX MAST catalog, the magnitude is 20.78 in the FUV channel, which means $2\times 10^{-5}$ Jy at the distance of the star from the Sun or
$2.7\times 10^9$ Jy at 1 au from GD 362.
Thus GD 362 is approximately 35\% as bright as the Sun (in the FUV), if viewed from the same distance.
That means the photodissociation time at 1 au from GD 362 would be 3 days.
At the inner and outer radii of the mid-infrared dust disk (\S\ref{sec:model}), the
dissociation times would be 10 to 800 sec. 
Thus the environment around GD 362 is quite hostile for water, which would be rapidly dissociated into OH,
which further dissociates to O and H in a timescale only slightly longer than H$_2$O. This explains why there are no narrow water vapor lines in the JWST spectrum.

\subsection{Hydrogen abundance in GD 362\label{sec:hydrogen}}
The atmosphere of GD 362 is dominated (by mass) by helium \citep{zuckermanChemicalCompositionExtrasolar2007};
however, there is definitely some hydrogen in the atmosphere
as evidenced by the Balmer and Paschen absorption lines seen in the
optical \citep{gianninasDiscoveryCoolMassive2004} 
and our new JWST  (Fig.~\ref{fig:nirspec}) spectrum.
\citet{koesterAccretionDiffusionWhite2009} calculated a  mass of H in the atmosphere of GD 362 of $7\times 10^{24}$ g. If this material is provided by external sources,
then a large accretion of H-bearing material is required.
While we see no strong evidence for hydrogen-bearing species in the 
debris disk around GD 362, we can use the upper limit on water ice to
place an upper limit of $10^{16}$ g on hydrogen within the dust 
cloud. Extrapolating to larger bodies with a Dohnanyi size distribution,
and using the largest allowed $M_{\rm dust}$ (from \S\ref{sec:mass}),
the limit on total hydrogen in the solids is $4\times 10^{22}$ g.
This upper limit is 2 orders of magnitude lower than the amount required to supply the atmospheric H.
It could be accommodated by changing the 
size distribution index to a much shallower $p=3.35$ and requiring an
Earth-sized largest body, but such a distribution is unlikely to be 
sustainable for a collisional cascade.

The upper limit on hydrogen in the debris disk is only within the context of the minerals included in the models used to fit the JWST spectra.
The limit on the mass of H from water relative to the mass of Si
is 0.4\% (twice the amount of H in H$_2$O that can marginally be included 
in the spectrum and obtain a good fit).
Hydrogen could also be present in other species. 
Water is the main carrier of H in chondrites, together with 
phyllosilicates and organics.
There is no evidence for phyllosilicates in the infrared spectra, where
their signature would be detectable.
To be generous, if 10\% of the silicates were misidentified and
actually were fully hydrated smectite, which has 12 H atoms per 4 Si,
the mass abundance ratio of H/Si would be 1\%.
For organics, a similar upper limit can be derived.
If all of the amorphous carbon were aromatic hydrocarbons, then
its H abundance (by mass) relative to Si, would be 3\% (1/12 of the
C/Si mass abundance).
However, even this upper limit on H from organics is too permissive,
as the carbonaceous material cannot be all organics. However, hydrocarbons have
strong features at 3.3--3.4 $\mu$m, which are seen in the JWST
spectra of planetary materials \citep[e.g.][]{woodwardJWSTStudyRemarkable2025} but are not
seen at all in our spectrum of GD 362 (Fig.~\ref{fig:nirspec}).
Thus the mass of H is $< 1$\% of the mass of Si, whether in
water, phyllosilicates, or organics.

The abundance of H in the GD 362 dust is in fact low compared to that seen in meteoritic 
 material.
From \citet{loddersRelativeAtomicSolar2021},
the abundance of 
H relative to Si in chondritic material is 9\% for CM chondrites
(17\% for CI chondrites), while the abundance of C relative to Si is
18\% for CM chondrites (38\% for CI chondrites).

\citet{koesterDBWhiteDwarfs2015} found that 75\% of DB white dwarfs with high signal-to-noise ratio spectra
show trace amounts of hydrogen in their atmospheres. Similarly, \citet{kilic100PcWhite2025} found traces of
hydrogen in 35 out of 50 DB white dwarfs in the SDSS 100 pc white dwarf sample. This fraction is significantly
higher than the fraction of metal-polluted white dwarfs in the solar neighborhood \citep{koesterFrequencyPlanetaryDebris2014},
suggesting a different origin for the prevalence of hydrogen in DB white dwarfs.
\citet{juraWaterFractionsExtrasolar2012} found that external pollution in the overall DB white dwarf population is $<1$\% by mass
composed of water, supporting the idea that external pollution may not be
the only source of hydrogen in DB atmospheres.

Modeling the spectral evolution of white dwarfs, \citet{bedardSpectralEvolutionHot2023} predicted the presence of a massive hydrogen reservoir underneath the
thin superficial layer \citep[see][]{rollandConvectiveDredgeupModel2020}, and demonstrated that the trace amounts of hydrogen in DBA white dwarfs in most cases is primordial.
However, external accretion of hydrogen surely happens in at least a  fraction of metal-polluted white dwarfs.
Interestingly, the five most H-rich stars among the helium-atmosphere white dwarfs all show evidence of metal accretion.
In addition, hydrogen is twice as common in metal-polluted helium-atmosphere white dwarfs compared to their metal-free counterparts
\citep{gentilefusilloTraceHydrogenHelium2017}. 
Hence, at least a fraction of the hydrogen seen in these objects is likely from water \citep[see also][]{juraSurvivalWaterExtrasolar2010,farihiEvidenceWaterRocky2013,raddiLikelyDetectionWaterrich2015}.

Our JWST spectrum as well as the photospheric abundances \citep{zuckermanChemicalCompositionExtrasolar2007,xuTWOBEYONDPRIMITIVEEXTRASOLAR2013} show
that the material accreting onto GD 362 is dry, and there is no evidence of significant hydrogen accretion from the disk currently orbiting GD 362.
However, the absence of evidence does not mean evidence of absence. GD 362 is clearly an outlier in its hydrogen abundance. Hydrogen
never diffuses out of the atmosphere in white dwarfs. Hence, the large amount of hydrogen in GD 362 may be  due to past accretion events where
water-rich minor bodies were ingested.

\section{Conclusions\label{sec:conclusion}}

The JWST mid-infrared spectrum of the polluted white dwarf GD 362
provides a significant increase
in spectral resolution and sensitivity, allowing us to measure the composition
of solid planetary material aroudn the star.
The mid-infrared emission can be explained by a disk that 
absorbs 2.4\% of the luminosity of the white dwarf and
emits throughout the mid-infrared. 
A recent JWST spectral (5--13 $\mu$m) survey of other white dwarfs showed a wide range of dust luminosity,
with the ratio of the 8--12 $\mu$m emission from the dust to that from the star
ranging from 0.3 to 36 \citep{farihiSubtleSpectacularDiverse2025}. That
same ratio for GD 362 is 82, indicating  how exceptionally dusty it is.
The GD 362 spectrum is similar to 6 of the surveyed stars, and it is most
similar in shape to the dustiest of the stars in that survey
(WD J1612+554). The amplitude of the silicate feature relative to 
the underlying continuum is three times greater for WD J1612+554
than for GD 36, indicating a wide variety of dust properties around different white dwarfs.

There is no evidence for a significant
variation of the brightness of the GD 362 debris since the prior observations
with Spitzer, though the data are consistent with a decrease
in brightness by $5\pm 3$\% over the 18 years between observations.

Though GD 362 is too distant to directly image any extrasolar planets 
around it,
the JWST images are deep enough to show that there are no
companion stars or giant planets larger than 25 times Jupiter's mass.

The mid-infrared spectrum is dominated by an exceptionally strong 9--11 $\mu$m silicate feature, which
is three times brighter than its underlying continuum. 
The continuum extends to at least 2 $\mu$m, and it requires hot debris (950 K) close to the star.
The emission from the disk, modeled with the Monte Carlo ray-tracing 
code {\it mcfost}, extends from just outside the Roche limit of the white dwarf to 0.08 au. Models with scale height larger than half the stellar radius
can fit the spectrum. Flatter disks cannot absorb enough starlight while
matching the continuum shape, nor do they produce the silicate feature.

A multi-mineral linear decomposition of the spectrum shows a mix of amorphous and crystalline olivines and pyroxenes plus amorphous
carbon. No phyllosilicates are found.
The stoichiometry of the detected minerals shows that C, O, Mg, Al, Fe 
are within a factor of 2 of chondritic abundances, relative to Si. 
There is no evidence for H$_2$O in the spectrum, nor were other
H-bearing species found, suggesting the dust
is drier and lower in hydrogen than in chondritic meteorites.

Overall the results indicate that GD 362 is surrounded by a disk with solids having elemental abundances
approximately matching those seen in the atmosphere of the white dwarf, supporting the connection between disk
and atmosphere arising from accretion of planetary material.

\begin{acknowledgments}
This work is based on observations made with the NASA/ESA/CSA James Webb Space Telescope. These observations are associated with program \#2919.
Support for program \#2919 was provided by NASA through a grant from the Space Telescope Science Institute, which is operated by the Association of Universities for Research in Astronomy, Inc., under NASA contract NAS 5-03127.
The data were obtained from the Mikulski Archive for Space Telescopes at the Space Telescope Science Institute. 

M.K. acknowledges support by the NSF under grants AST-2205736 and AST-2508429, and by and NASA under grant Nos. 80NSSC22K0479, 80NSSC24K0380, and 80NSSC24K0436.
\end{acknowledgments}

\facility{JWST,Spitzer} 
\def\extra{
\software{
astropy \citep{astropycollaboration2022},
matplotlib \citep{hunterMatplotlib2DGraphics2007}
}
}
\def\apj{ApJ}

\bibliography{references}{}
\bibliographystyle{aasjournalv7}

\end{document}